\documentclass[journal]{IEEEtran}

\usepackage{subfigure}
\usepackage{amsmath}
\usepackage{amsfonts}
\usepackage{amssymb}
\usepackage{mathrsfs}
\usepackage{dsfont}
\usepackage{bbm}
\usepackage{bm}
\usepackage{color}
\usepackage{hyperref}
\usepackage{booktabs}
\usepackage{graphicx}
\usepackage{epstopdf}
\usepackage{epsfig}
\usepackage{cite}
\usepackage{enumerate}
\usepackage{amsopn}
\usepackage{ntheorem}
\usepackage{algorithm,algpseudocode}

\newtheorem{theorem}{Theorem}

\newtheorem{lemma}{Lemma}

\newtheorem{definition}{Definition}
\newtheorem{problem}{Problem}
\newtheorem{assumption}{Assumption}

\DeclareMathOperator{\Tr}{Tr}
\DeclareMathOperator*{\argmin}{arg\,min}

\makeatletter

\newcommand{\Rmnum}[1]{\expandafter\@slowromancap\romannumeral #1@}
\makeatother

\title{
Optimal Scheduling of Multiple Sensors over Lossy and Bandwidth Limited Channels
}

\author{Shuang Wu$^*$, Kemi Ding$^\dagger$, Peng Cheng$^\ddagger$, Ling Shi$^*$
\thanks{$*$: Electronic and Computer Engineering, Hong Kong University of Science and Technology, Clear Water Bay, Kowloon, Hong Kong, (e-mail: {swuak@ust.hk}, {eesling@ust.hk}).}
\thanks{$^\dagger$: School of Electrical, Computer and Energy Engineering, Arizona State University, Tempe, USA (e-mail: {kding11@asu.edu}).}
\thanks{$^\ddagger$: State Key Laboratory of Industrial Control Technology, Zhejiang University, Hangzhou, China (e-mail: {pcheng@iipc.zju.edu.cn}).}
\thanks{The work by S. Wu and L. Shi is supported by a Hong Kong RGC General Research Fund 16204218.}
\thanks{The work by P. Cheng was partially supported by NSFC under grant 61761136012 and 61533015.}
}

\begin{document}
\maketitle

\begin{abstract}
This work considers the sensor scheduling for multiple dynamic processes. We consider $n$ linear dynamic processes. The state of each process is measured by a sensor, which transmits its local state estimate over one wireless channel to a remote estimator with certain communication costs. At each time step, only a portion of the sensors are allowed to transmit data to the remote estimator and the packet might be lost due to unreliability of the wireless channels. Our goal is to find a scheduling policy which coordinates the sensors in a centralized manner to minimize the total expected estimation error of the remote estimator and the communication costs. We formulate the problem as a Markov decision process. We develop an algorithm to check whether there exists a deterministic stationary optimal policy. We show the optimality of monotone policies, which saves the computational effort of finding an optimal policy and facilitates practical implementation. Nevertheless, obtaining an exact optimal policy still suffers from curse of dimensionality when the number of processes is large. We further provide an index-based heuristic to avoid brute force computation. We derive analytic expressions of the indices and show that this heuristic is asymptotically optimal. Numerical examples are presented to illustrate the theoretical results.
\end{abstract}

\begin{IEEEkeywords}
Kalman filtering; Sensor scheduling; lossy network; monotone policy; Markov decision process; index policy 
\end{IEEEkeywords}
\section{INTRODUCTION}
The development of device, sensing and communication technologies enables wide range of applications of wireless sensor networks. After the pioneering work on event-based sensor data scheduling proposed in~\cite{astrom2002comparison}, a variety of studies has been done to balance the estimation performance and the communication overhead in~\cite{imer2005optimal,leong2017sensor,ren2017infinite}.

A large number of works on sensor scheduling focused on remote estimation of a linear time-invariant (LTI) dynamic process. There are also some other works addressing static processes and nonlinear models. However, the static models~\cite{gao2018optimal,vasconcelos2018optimal} are special cases of LTI systems and nonlinear models either involve approximation of a linear system~\cite{lin2009energy,shuman2010measurement} or the solution method requires numerically solving a partially observable Markov decision process, which is computationally inefficient~\cite{krishnamurthy2002algorithms,he2004sensor,krishnamurthy2007structured}. A few works~\cite{molin2014price,gatsis2015opportunistic} considered control problems with transmission constraints, which can also be transformed into sensor scheduling problems as they prove the separation between optimal controls and optimal transmissions.

The sensor scheduling problems have been modeled in different frameworks. A number of works modeled it as a Markov decision process (MDP), which is a framework for optimal stochastic control problems. Obtaining an optimal solution of an MDP involves stochastic dynamic programming-based numerical algorithms such as a value iteration and a policy iteration, which prohibits solving large-scale problems due to the curse of dimensionality. Therefore, most works only use MDP to deal with a single process~\cite{nayyar2013optimal,akyol2014controlled,leong2017sensor}. When there is only one dynamic process, an approximation of the optimal sensor scheduling policy can also be obtained by analyzing a modified algebraic Riccati equation (MARE), which characterizes the dynamics of the remote estimation error. Zhao et al.~\cite{zhao2014optimal} studied the asymptotic behavior of the MARE and showed that the optimal policy can be approximated by a periodic one. Orihuela et al.~\cite{orihuela2014periodicity} further showed that a periodic policy is optimal under a myopic criterion. Some other works modeled the sensor scheduling problem as static sensor selection problems, resulting in an optimization problem in an Euclidean space with integer constraints. They either found a convex approximation of the original problem~\cite{mo2011sensor} or used some greedy based heuristics to find a suboptimal policy with theoretical performance bound~\cite{asghar2017complete}. Although efficient algorithms can be developed from approximated models, the gap between the approximated policy and the optimal policy can be significant.

The framework for a sensor scheduling problem depends on the information available for scheduling. If there is only offline information, such as system parameters, open loop scheduling is enough. The sensors transmit data based on system clock and predetermined timing. The periodic policy~\cite{zhao2014optimal,orihuela2014periodicity} and static sensor selection~\cite{mo2011sensor,asghar2017complete} aforementioned are in this category. Besides offline scheduling, a large number of works were devoted to optimal online scheduling. Since additional online information is available, an online scheduling policy may yield better performance than an offline one. Nevertheless, analysis and design an online policy is nontrivial.

Online information can be further categorized into two classes: system state information and holding time information. System state information refers to the actual system state if the state is fully observable, or the innovation of the measurements if the state observation is noisy. Once the size of the system state is greater than a threshold value, a sensor will be scheduled to transmit data. Therefore, these scheduling policies are also termed as data-driven or event-based. Works on data-driven scheduling mostly focus on the single sensor case~\cite{molin2012optimality,wu2013event,han2015stochastic,shi2016event}. Scheduling of multiple sensor with the system state poses significant challenges in light of coordination. Xia et al.~\cite{xia2016networked} showed that, if no coordination of the sensor transmissions is considered, the potential transmission collisions will cause an online policy to perform worse than an offline policy. Molin and Hirche~\cite{molin2014price} considered LQG control with fully observable states of multiple systems under a communication rate budget, which is inapplicable if the number of allowable channels is limited at every time step. Gatsis et al.~\cite{gatsis2015opportunistic} considered transmission power minimization under a system stability constraint. This cannot be applied if we aim to minimize the estimation error. Holding time information is the time elapsed since the remote estimator receives data from the sensors. In telecommunication society, this concept attracts a growing interest and is termed as the age of information (AoI)~\cite{kadota2018optimizing}. In this work, we shall see that there is a one-to-one correspondence between the holding time information and system performance if the sensors are able to conduct local computations. This facilitates design and analysis as the holding time only takes values in the set of positive integers. Leong et al.~\cite{leong2017sensor} utilized this property to study the optimal scheduling for one dynamic system over a lossy channel. If there is no packet dropout in the communication channel, the holding time becomes offline information as the packet arrival sequence is available before actual transmissions. In this case, the online problem is reduced to the offline one.

In this work, we consider multiple sensor scheduling using online holding time information of multiple dynamic processes, which is an extension of previous works~\cite{shi2012scheduling,han2017optimal}. In these works, only unstable processes over a reliable channel were considered. We generalize the results to a setup where both stable and unstable processes exist over lossy channels. We use MDP to formulate the problem. Although the framework has been studied, the analysis fails to work for stable processes as mentioned in~\cite{han2017optimal}. If there are no packet dropouts, the state space can be restricted to be finite as done by~\cite{han2017optimal}. If the channel is lossy, however, the existing approach of~\cite{han2017optimal} no longer works. In addition, we take the costs of communication into consideration, which has not been addressed previously since the one-stage cost becomes more complicated. We show the optimality of a monotone deterministic stationary policy. Furthermore, we use the celebrated Whittle's index~\cite{whittle1988restless} to develop a heuristic policy, which can be written in a closed-form and is asymptotically optimal.

The contribution of our work is multi-fold.

(1) We develop an algorithm-based sufficient condition for existence of a deterministic stationary optimal policy, which generalizes the approaches in previous works (e.g.,~\cite{shi2012scheduling,han2017optimal}). We formulate the multi-sensor scheduling problem as an average cost Markov decision process (MDP) over an infinite horizon. As the communication channel is lossy, the state space of an MDP over an infinite horizon is infinite and there may not be an optimal policy in the class of deterministic stationary policies. We develop \textbf{Algorithm~\ref{alg:feasibility}} and show that deterministic stationary optimal policies indeed exist if the output of the algorithm is greater than the number of available channels.

(2) We show the optimality of monotone policies (\textbf{Theorem~\ref{theorem: monotonicity for multiple processes}}), which sheds light on the structure of optimal policies. In particular, if it is optimal to schedule a sensor in one state, it is also optimal to schedule this sensor when the state of this sensor increases while others remaining unchanged. Although dynamic programming can be used as a general approach to tackle MDPs, only numerical solutions can be obtained and no design insights of an optimal policy can be acquired. The monotone structure seems intuitive, but its proof is not straightforward.

(3) We use the Whittle's index~\cite{whittle1988restless} to develop an index-based heuristics for the scheduling policy (\textbf{Theorem~\ref{theorem: whittle's index}}) instead of solving the problem via brute-force numerical algorithms. The index-based policy provides an asymptotically optimal policy without using brute force numerical algorithms to solve the MDP. Although such heuristics have been adopted in several problems in an MDP setup, e.g.,~\cite{liu2010indexability,larranaga2015stochastic,borkar2018opportunistic,kadota2018optimizing} computing the Whittle's index generally requires an iterative algorithm. We derive analytic expression of these indices in this work, which reduces computation overhead significantly and facilitates online implementation.

The remainder of this paper is organized as follows. In section~\Rmnum{2}, we present the mathematical formulation of the problem of interest. In section~\Rmnum{3}, we present the MDP formulation and the optimality of a monotone deterministic stationary policy. In section~\Rmnum{4}, we construct a Whittle's index-based suboptimal heuristics. The numerical examples in section~\Rmnum{5} are provided to demonstrate the monotone policies and performance of the index-based policy. We summarize the paper in section~\Rmnum{6}. We leave all proofs in the Appendix.

\emph{Notation}: Denote $\mathbb{N}$ and $\mathbb{R}$ as the set of nonnegative integer numbers and real numbers, respectively. The symbol $\mathbb{X}^n$ stands for the $n$-th order Cartesian product of a set $\mathbb{X}$. Inequalities (i.e., $<,>,\leq,\geq$) between two vectors are interpreted an element-wise. For a matrix $X$, let $\Tr(X)$, $X^{\top}$ and $\rho(X)$ represent the trace, the transpose and the spectral radius of $X$, respectively. The symbol $I$ stands for an identity matrix of appropriate size. Let $\mathtt{Pr}(\cdot)$ and $\mathtt{Pr}(\cdot|\cdot)$ stand for the probability and conditional probability for certain events. Denote $\mathbb{E}[\cdot]$ as the expectation of a random variable. The composition of two mappings $f$ and $g$ is denoted by $g\circ f$ and the composition of a mapping $f$ for $t$ times is denoted by $f^t:=\underbrace{f \circ f \circ \cdots \circ f}_t$ with $f^0$ being the identity mapping. A Lyapunov operator is defined as $h_i(X):= A_iXA_i^{\top}+Q_i$.

\section{SYSTEM SETUP AND PROBLEM FORMULATION}
\subsection{System Setup}
\begin{figure}[t]
	\centering
    \includegraphics[width=0.4\textwidth]{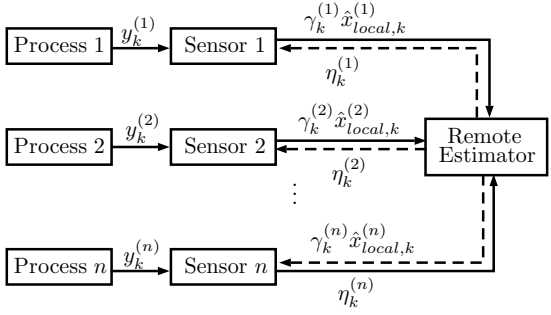}
    \caption{Architecture of the remote estimation system.}
    \label{fig:architecture}
\end{figure}

Consider the remote estimation system in Fig.~\ref{fig:architecture}. We illustrate each component as follows.

\emph{Processes}. There are $n$ independent discrete-time linear dynamic processes whose states are measured by $n$ sensors, respectively. This type of system configuration can be implemented with the \emph{Wireless}HART protocol in industrial applications~\cite{song2008wirelesshart}. The dynamics of the sensor system is as follows:
$$x^{(i)}_{k+1} = A_ix^{(i)}_{k} + w^{(i)}_k,~y^{(i)}_k = C_i{x}^{(i)}_{k} + {v}^{(i)}_k,$$
where $i \in \{1,\ldots,n\}$, ${x}_k^{(i)}\in\mathbb{R}^{n_i}$ is the state of the $i$-th system at time $k$ and ${y}^{(i)}_k\in\mathbb{R}^{m_i}$ is the noisy measurement taken by sensors. For all processes and $k\geq0$, the state disturbance noise ${w}_k^{(i)}$, the measurement noise ${v}_k^{(i)}$ and the initial state ${x}_0^{(i)}$ are mutually independent Gaussian random variables, which follow Gaussian distributions as ${w}_k^{(i)}\sim\mathcal{N}({0},{Q}_i)$, ${v}_k^{(i)}\sim\mathcal{N}({0},{R}_i)$ and ${x}_0^{(i)}\sim\mathcal{N}({0}, {\Sigma}^x_i)$. We assume that ${Q}_i$ and ${\Sigma}^x_i$ are positive semidefinite, and ${R}_i$ is positive definite. We assume that, for every $i \in N$, the pair $({A}_i,{C}_i)$ is detectable and the pair $({A}_i,\sqrt{{Q}}_i)$ is stabilizable.

\emph{Sensors}. Each sensor is assumed to be equipped with computation unit and memory capacity. After taking the measurement, the sensor computes $\hat{{x}}_{local,k}$, the local minimum mean squared error estimate of the state ${x}^{(i)}_k$ at each time step based on the Kalman filter~\cite{kalman1960new}. After computation, the sensor transmit the local state estimates if the remote estimator delivers a transmission order to it through a feedback channel.


\emph{Communication channels}. The communication bandwidth is considered to be limited. At each time step, the remote estimator can only receive data from $m$ out of the $n$ sensors through a forward channel. Let $a^{(i)}_k \in \{0, 1\}$ denote whether the $i$-th sensor is scheduled to transmit data at time $k$. This command is sent from the remote estimator to the sensor through the feedback channel. If the remote estimator decides to ask for data of sensor $i$ at time $k$, $a_k^{(i)}=1$; otherwise, $a_k^{(i)}=0$.
We also consider the unreliability of the channel. Let $\eta_k^{(i)} \in \{0, 1\}$ denote whether the packet is successfully received by the remote estimator through the forward channel. Let $\eta^{(i)}_k=1$ stand for successful transmission, and $\eta^{(i)}_k=0$ for failure. Similar to the setting in~\cite{ren2018attack}, the channel condition is assumed to be independently distributed and $\mathbb{E}[\eta^{(i)}_k]=\lambda_i$, for any $k\geq0$. For the feedback channel, similar to other references in the literature~\cite{mo2014detecting}, the transmission is assumed to be reliable since the remote estimator is typically able to transmit signal with greater power.

\emph{Remote estimator}. Let the random variable $\xi^{(i)}_k=a^{(i)}_k\eta_k^{(i)}$ denote whether a local estimate of sensor $i$ is received by the remote estimator. According to~\cite{anderson1979optimal}, since $({A}_i,{C}_i)$ are detectable and $({A}_i,\sqrt{{Q}_i})$ are stabilizable, the \emph{a posteriori} estimation error covariance ${P}_{local,k}^{(i)}$ converges exponentially fast to a steady state $\overline{{P}}^{(i)}$, usually in a few steps. We assume that the system operates in the steady state. Based on this fact, the optimal estimate of each process for the remote estimator is as follows:
\begin{align*}
\hat{{x}}_k^{(i)}=
\begin{cases}
\hat{{x}}^{(i)}_{local,k}, &\text{if~} \xi_k^{(i)}=1,\\
{A}_i \hat{{x}}^{(i)}_{k-1}, &\text{if~} \xi_k^{(i)}=0.
\end{cases}
\end{align*}
Define the time elapsed since the last received packet of the $i$-th sensor at time $k$:
\begin{align}\label{eq:definition of tau}
\tau^{(i)}_k=\min_{t}\{0\leq t\leq k:\xi^{(i)}_{k-t}=1\}.
\end{align}
The estimation error covariance matrices at the remote estimator are thus as follows:
\begin{align*}
{P}_k^{(i)}=
\begin{cases}\overline{{P}}^{(i)}, &\text{if~} \xi_k^{(i)}=1,\\h_i({P}_{k-1}^{(i)}), &\text{if~} \xi_k^{(i)}=0.\end{cases}
\end{align*}
The estimation error covariance of the remote estimator can be compactly written as
\begin{align}
P_k^{(i)}=h_i^{\tau_k^{(i)}}(\overline{P}^{(i)}).\label{eq:remote P}
\end{align}
According to~\cite[Lemma 3.1]{shi2012scheduling}, the operator $h_i^{\ell}({X})$ is monotonically increasing with respect to $\ell$, i.e., $\forall i \in N$, if $\ell_1\leq\ell_2$ for $\ell_1,~\ell_2 \in \mathbb{N}$, $h_i^{\ell_1}(\overline{{P}}^{(i)})\leq h_i^{\ell_2}(\overline{{P}}^{(i)})$. Moreover, $\forall \ell \in \mathbb{Z}_+$, $\Tr(\overline{{P}}^{(i)})<\Tr(h(\overline{{P}}^{(i)}))<\cdots<\Tr(h^{\ell}(\overline{{P}}^{(i)}))$.

\subsection{Problem Formulation}
From~\eqref{eq:remote P}, the expected estimation error covariance is a function of ${\tau^{(i)}_k}$ and is independent of the realization of $\hat{x}^{(i)}_{local,k}$. As the remote estimation error covariance now has a one-to-one correspondence with $\tau_k^{(i)}$, we denote the cost associated with the remote estimation error as
\begin{align*}
c_e^{(i)}(\tau^{(i)}_k)=\Tr(P_k^{(i)}).
\end{align*}
We also take energy consumption of the sensors into consideration. If sensor $i$ transmit data, an energy cost $c_c^{(i)}$ is incurred for sensor $i$. Our objective is to find a scheduling policy $\{a_k^{(i)}:i=1,2,\dots,n;\;k=0,1,2,\dots\}$ to minimize the expected time-averaged trace of the remote estimation error and the normalized energy cost over all sensors as follows.

\begin{problem}\label{prb:problem1}
	\begin{align*}
	\min_{\{a_k^{(i)}\}} \quad &\lim_{T\to \infty} \frac{1}{T+1} \sum_{k=0}^{T} \sum_{i=1}^{n} \mathbb{E} [c_e^{(i)}(\tau_k^{(i)})+  c_c^{(i)}a_k^{(i)}]\\
	\text{s.t.} \quad &\sum_{i=1}^n a_k^{(i)} \leq m, ~\forall k\geq0.
	\end{align*}
\end{problem}

The feasibility of Problem \ref{prb:problem1} requires that there exists a policy such that the objective function is bounded. A necessary condition is imposed as follows.
\begin{assumption}\label{assumption:neccessary for stability}
$
\max_i \rho^2(A_i)(1-\lambda_i) < 1.
$
\end{assumption}
This assumption ensures that the estimation error covariance of each process is bounded if every sensor is allowed to transmit simultaneously at each time step. This assumption is only a necessary condition to ensure the existence of a solution to the problem as the constraint on the number of simultaneous sensor transmissions is neglected. We develop a sufficient condition in Theorem~\ref{thm:existence} in the next section.

\section{Structural Properties of an Optimal Policy}
In this section, we first formulate Problem~\ref{prb:problem1} as a Markov decision process (MDP) with average cost over an infinite horizon. We then present an algorithm-based sufficient condition to guarantee the existence of a deterministic stationary optimal policy for the MDP. We show that there exist monotone structures in an optimal stationary policy, which extends the threshold structure of single sensor scheduling to a multiple-sensor case.

\subsection{MDP Formulation}
The form of \textbf{Problem~\ref{prb:problem1}} can be taken as an MDP with an infinite time-averaged cost which consists of a quadruple $(\mathbb{S},\mathbb{A},\mathtt{Pr}(\cdot|\cdot,\cdot),c(\cdot,\cdot))$. Each element is explained as follows.

1) The state space $\mathbb{S}$ contains all possible states ${s} := [\tau^{(1)},\ldots,\tau^{(n)}]^\top \in \mathbb{N}^n$, where $\tau^{(i)}$ is a shorthand notation for $\tau^{(i)}_k$ defined in~\eqref{eq:definition of tau} by omitting the time index $k$. This can be done because we are going to discuss the transition between two successive time steps, where the time index $k$ is not necessary.

2) The action space $\mathbb{A}$ contains all allowable scheduling actions, i.e., $\mathbb{A} := \{a = [a^{(1)},\dots,a^{(n)}]\in\{0,1\}^n:a^{(i)}\in\{0,1\}, \forall i=1,\ldots,n, \sum_{i=1}^na^{(i)}\leq m\}$, where $a^{(i)}=1$ stands for scheduling the $i$-th sensor and $0$ otherwise.

3) At time $k$, suppose the state is in $s_k=s$. After taking action $a_k=a$, the state will transit to another state $s_+$ in the next time step by following a time-homogeneous transition law as follows.
\begin{align}
\mathtt{Pr}(s_+|s,a)=\prod_{i=1}^n \mathtt{Pr}^{(i)}(\tau^{(i)}_+|\tau^{(i)},a^{(i)}),
\end{align}\label{eq:transition law}
where
\begin{align}\label{eq:single process transisiton}
\mathtt{Pr}^{(i)}(\tau^{(i)}_+|\tau^{(i)},a^{(i)})=
\begin{cases}
\lambda_i, &\text{if~} \tau^{(i)}_+=0,a^{(i)}=1,\\
1-\lambda_i, &\text{if~} \tau^{(i)}_+=\tau^{(i)}+1,a^{(i)}=1,\\
1, &\text{if~} \tau^{(i)}_+=\tau^{(i)}+1,a^{(i)}=0,\\
0, &\text{otherwise.}
\end{cases}
\end{align}

4) The one-stage cost is defined as
$c({s},{a}) := \sum_{i=1}^n c_e^{(i)}(\tau^{(i)})+  c_c^{(i)}a^{(i)}$.

Let $(s_{0:k},a_{0:k-1})=({s}_0,{a}_0,\dots,{s}_{k-1},{a}_{k-1},{s}_k)$ stand for the history up to time $k$. A policy is a sequence of mappings from the history to a probability distribution of the scheduling actions, i.e., $\{\pi_k\}_{k=0}^\infty$, where  $\pi_k:(s_{0:k},a_{0:k-1}) \mapsto \mathtt{Pr}(a_k)$. Let $\Pi$ denote the set of all feasible policies. The goal of an MDP is to minimize the expectation of a time-averaged cost over an infinite horizon as
\begin{align*}
\min_{\{\pi_k\}_{k=0}^\infty\in\Pi} \lim_{T\to \infty} \frac{1}{T+1} \sum_{k=0}^{T} \sum_{i=1}^{n} \mathbb{E} [c_e^{(i)}(\tau_k^{(i)})+  c_c^{(i)}a_k^{(i)}].
\end{align*}

\subsection{Existence of Deterministic Stationary Policy}
The general policy class $\Pi$ requires the information of the whole history and could be random, which hinders practical scheduling implementations. In this work, we consider deterministic stationary policies with the form
\begin{align*}
a_k = \pi(s_k)
\end{align*}
where $\pi=\pi_k$ for any $k\geq0$. These policies are more desirable, as the actions are deterministic and the mappings are stationary (independent of time $k$).

We introduce Algorithm~\ref{alg:feasibility}, the output of which determines whether optimal policies can be found in the set of deterministic stationary ones. Let $G^{(u)}:=\{G^{(u)}[i]:\rho(A_{G^{(u)}[i]})\geq 1\}$ be the set of the indices of all unstable processes. Given the necessary condition (Assumption~\ref{assumption:neccessary for stability}), Algorithm~\ref{alg:feasibility} gives the least number of channels such that all the processes are stabilizable.

\begin{algorithm}
	\caption{Feasibility of Multiple Sensor Scheduling}
	\label{alg:feasibility}
	\begin{algorithmic}[1]
		\State Initialize the group number counter $\Bbbk\leftarrow 1$ and the first group $G_1\leftarrow\{G^{(u)}[1]\}$
		\For{Process $i=G^{(u)}[2]:|G^{(u)}|$}
		\For{$j=1:\Bbbk$}
		\If{Process $i$ and process in Group $G_j$ satisfy
			\begin{align*}
			\max_{i'\in G_j \bigcup\{i\}}\rho^2(A_{i'})\max_{j'\in G_j \bigcup\{i\}}(1-\lambda_{j'})<1
			\end{align*}
			\hspace{2.5em}}
		\State $G_j\leftarrow G_j \bigcup\{i\}$ and \textbf{break}
		\EndIf
		\EndFor
		\If{process $i$ has not been put in any group}
		\State{$\Bbbk\leftarrow \Bbbk+1,~{\Bbbk}\leftarrow\{G^{(u)}[i]\}$}
		\EndIf
		\EndFor
		\State Output $\Bbbk$
	\end{algorithmic}
\end{algorithm}

The following theorem characterizes a sufficient condition for existence of a deterministic stationary optimal policy for the MDP formulation.

\begin{theorem}\label{thm:existence}
	If the output in Algorithm~\ref{alg:feasibility} is less than or equal to $m$,
	there exist a constant $\mathcal{J}^\star$, a function $V^\star(\tau)$, and a deterministic stationary policy $\pi^\star:\mathbb{S}\mapsto\mathbb{A}$ that satisfy the following Bellman optimality equation
	\begin{align*}
	\mathcal{J}^\star + V^\star(s) = \min_{a\in\mathbb{A}} \Bigg[ c(s,a) + \sum_{s_+\in\mathbb{S}} V^\star(s_+)\mathtt{Pr}(s_+|s,a) \Bigg]
	\end{align*}
	and
	\begin{align*}
	\mathcal{J}^\star + V^\star(s) = \Bigg[ c(s,\pi^\star(s)) + \sum_{\tau_+\in\mathbb{S}} V^\star(s_+)\mathtt{Pr}(s_+|s,\pi^\star(s)) \Bigg].
	\end{align*}
	In addition,
	\begin{align*}
	J(\pi^\star) = \min_{\pi\in\Pi} J(\pi) = \mathcal{J}^\star.
	\end{align*}
\end{theorem}

This theorem shows that it is nontrivial to establish the existence of a regular optimal policy for the multiple sensor scheduling problem if packet dropouts occur. Roughly speaking, if the channel bandwidth is sufficient, there exists a deterministic stationary optimal policy. In previous works~\cite{shi2012scheduling,han2017optimal} on scheduling of multiple linear dynamic processes, a perfect channel is assumed. Our problem, however, considers a lossy channel. As a result, the number of the feasible consecutive packet loss cannot be restricted to be finite as it was done in~\cite{han2017optimal}. Therefore, proving the existence of a deterministic stationary policy is challenging. Furthermore, our result holds when there are stable processes. This extends the results of~\cite{han2017optimal}, which only considered unstable processes and cannot be extended to stable processes.

\subsection{Structure of an Optimal Policy}
One can directly obtain an optimal policy through relative value iteration or policy iteration for~\eqref{eq:acoe}. This, however, cannot provide more insights of the structure of the problem. One can observe that the one-stage cost $c(s,a)$ and the state transition law possesses certain monotone structure, which, leads to optimality of monotone policies.

\begin{theorem}\label{theorem: monotonicity for multiple processes}
There exists an optimal deterministic stationary policy $\pi^\star$ with a monotone structure. In particular, if $\tau^{(i)} \leq \tau'^{(i)}$ with $\tau^{(j)}=\tau'^{(j)}$ for $j  \neq i$ and the $i$-th component of $\pi^\star(\tau)$ is one, then the $i$-th component of $\pi^\star(\tau')$ is also one.
\end{theorem}

This result shows that, if it is optimal to schedule sensor $i$ at state $s$, it is also optimal to schedule sensor $i$ at state $s'$, where $\tau^{(i)}\leq\tau'^{(i)}$ and $\tau^{(j)}=\tau'^{(j)}$ for $j\neq i$. In particular, if $m=1$ and $n=2$, there exists a switching curve between scheduling or not scheduling one sensor in the state space. Examples can be found in the numerical example section.

The benefits of the monotone structure of the optimal policy are two-fold. Firstly, the structure policy reduces the storage space for online implementation. After obtaining the optimal scheduling policy, only the boundary state is needed to be stored for policy implementation. Secondly, by leveraging the monotone structure, we can reduce computation overhead of solving~\eqref{eq:acoe} compared with brute force numerical schemes such as relative value iteration or policy iteration. Following the idea in~\cite{zhou2017optimal}, the standard relative iteration can be revised as follows. The original relative value iteration iterates between the following two updates
\begin{align}
V_{k+1}(s) &= \min_{a\in\mathbb{A}} \Bigg[ c(s,a) + \sum_{s_+\in\mathbb{S}} V_k(s_+)\mathtt{Pr}(s_+|s,a) \Bigg],\label{eq:value update}\\
V_{k+1}(s) &= V_{k+1}(s) - V_{k+1}(s_o)\nonumber,
\end{align}
where $s_o\in\mathbb{S}$ is a fixed state. For each $k$, we can associate an optimal policy policy by letting
\begin{align}
\pi^{\star}_k(s)=\argmin_{a\in\mathbb{A}} \Bigg[ c(s,a) + \sum_{s_+\in\mathbb{S}} V_k(s_+)\mathtt{Pr}(s_+|s,a) \Bigg]\label{eq:policy improvement}
\end{align}
for each state $s$. In the revised version, before we compute~\eqref{eq:value update}, instead of minimizing for all state $s\in\mathbb{S}$, we first check whether there are $s'\leq s$ and $a\in\mathbb{A}$ such that $\pi^{\star}_k(s')=a$, and then let
\begin{align*}
\pi^{\star}_{k+1}(s) &= a,\\
V_{k+1}(s) &= c(s,a) + \sum_{s_+\in\mathbb{S}} V_k(s_+)\mathtt{Pr}(s_+|s,a)
\end{align*}
for the state $s$, if such $s'$ and $a$ exists. If such $s'$ and $a$ fail to exist, we execute the original update~\eqref{eq:value update} for $s$ and calculate $\pi^{\star}_{k+1}(s)$ via~\eqref{eq:policy improvement}. This revision removes the brute-force search over $\mathbb{A}$ in~\eqref{eq:policy improvement} by leveraging the monotone structure. According to Theorem~\ref{theorem: monotonicity for multiple processes}, the revised algorithm converges to the same policy as the original one. Similar revision can also be done for policy iteration. Details can be found in~\cite{zhou2017optimal}.

Scheduling multiple sensors is complex by its nature. When $n$ is large, storing the switching boundaries in $n$-dimensions is still intense. Moreover, although searching space of the relative value iteration and policy iteration has been reduced, the computation complexity is still exponential in $n$. In the next section, we present an index-based heuristics for the scheduling policy to further reduce computation overhead and to simplify the scheduling decisions.

\section{Index-Based Heuristics}
To obtain the optimal solution of the MDP, one needs to resort to a dynamic-programming-based numerical algorithm. Suppose that each process is approximated by $N$ states. There are $N^n$ states in total, which grows exponentially as $n$ increases. Meanwhile, the action space is $\sum_{i=0}^m\binom{n}{i}$. The large state space and action space make the brute force numerical methods prohibitive.

We construct an index-type heuristics based on the Whittle's index~\cite{whittle1988restless} to obtain a suboptimal scheduling policy. The index policy maps the each state of a sensor to a real number and determines which sensor to transmit based on the order of these real numbers. The mapping is calculated for sensors separately, which significantly reduces computation overhead.

As mentioned in Whittle's seminal paper~\cite{whittle1988restless}, several conditions are needed to ensure that the index policy can be constructed, which are known as indexability. The indexability requires case-by-case analysis. Generally, computation of the indices raises a significant challenge. Researchers use ad hoc approaches to tackle specific problems. We show that the index of the sensor scheduling in this model can be written in closed-form, which makes the index easy to compute and facilitates online implementation. In addition, this suboptimal policy is asymptotically optimal as the number of sensors and channels goes to infinity.

\subsection{Overview of the Index policy}
The derivation of the Whittle's index is based on regularization, which relaxes the hard constraint on simultaneous transmissions at each time step. This leads to decoupled sensor scheduling problems. We schedule sensors with the top $m$ largest indices if these indices are positive. Therefore, the actual index policy will still meet the hard constraint.

We start the analysis by transforming the hard constraint in Problem~\ref{prb:problem1}
\begin{align*}
\sum_{i=1}^n a_k^{(i)} \leq m, ~\forall k\geq0
\end{align*}
into a relaxed time-averaged form as
\begin{align}\label{eq: relaxed constraint}
\lim_{T\to \infty} \frac{1}{T+1} \sum_{k=0}^{T} \sum_{i=1}^{n} \mathbb{E} [a_k^{(i)}] \leq m.
\end{align}
We transform Problem~\ref{prb:problem1} into an unconstrained one by incorporating relaxed constraint in the objective functional with an extra penalty for transmission $w$, i.e.,
\begin{align*}
\min_\pi \lim_{T\to \infty} \frac{1}{T+1} \sum_{k=0}^{T} \sum_{i=1}^{n} \mathbb{E} [c_e^{(i)}(\tau_k^{(i)})+  c_c^{(i)}a_k^{(i)} + w a_k^{(i)}].
\end{align*}
This problem has a separable structure which can be further decoupled into $n$ independent scheduling problems
\begin{align}\label{eq:decoupled mdp}
\min_{\pi_i} \lim_{T\to \infty} \frac{1}{T+1}  \sum_{k=0}^{T} \mathbb{E} [c_e^{(i)}(\tau_k^{(i)})+  c_c^{(i)}a_k^{(i)} + w_i a_k^{(i)}]
\end{align}
for each $i$. This leads to $n$ \emph{decoupled MDPs}. Note that we further relax $w$ to $w_i$ for each $i$. By using the MDP framework in the last section, we have $n$ independent MDPs $(\mathbb{S}_i,\mathbb{A}_i,\mathtt{Pr}(\cdot|\cdot,\cdot),c^{(i)}(\tau^{(i)},a^{(i)}))$ with $c^{(i)}(\tau^{(i)},a^{(i)})=c_e^{(i)}(\tau^{(i)}) + c_c^{(i)}a^{(i)} + w_i a^{(i)}$, and the optimal policy for each $i$ can be characterized by the following Bellman optimality equation
\begin{align}
\mathcal{J}_i^\star + &V_i^\star(\tau^{(i)}) = \min_{a^{(i)}\in\mathbb{A}_i} \Bigg[ c_e^{(i)}(\tau^{(i)}) + c_c^{(i)}a^{(i)} + w_i a^{(i)}\nonumber\\ &+ \sum_{\tau^{(i)}_+\in\mathbb{S}_i} V_i^\star(\tau_+)\mathtt{Pr}^{(i)}(\tau^{(i)}_+|\tau^{(i)},a^{(i)}) \Bigg]\label{eq:acoe for each}.
\end{align}
An optimal policy determines whether $a^{(i)}=1$ or $a^{(i)}=0$ for each state $\tau^{(i)}$ and varies for different $w$. For each given state $\tau^{(i)}$, there exists a $w_i(\tau^{(i)})$ such that both $a^{(i)}=1$ and $a^{(i)}=0$ minimize the term inside the bracket in the right hand side of~\eqref{eq:acoe for each}. We can thus interpret $w_i(\tau^{(i)})$ as the importance of $\tau^{(i)}$. Whittle calls these $w_i(\tau^{(i)})$ indices. Whittle's original index policy runs as follows. Suppose that, for each $i$, the corresponding process is \emph{indexable} (see more details later). At each time step, we first \emph{sort} the index of each sensor according to their current state $\tau^{(i)}$ and then \emph{schedule} the $m$ sensors with largest indices.

\subsection{Derivation of the Index policy}
The key component of adopting the index policy is computing Whittle's index. Generally, this is computationally intense as the index $w_i(\tau^{(i)})$ is coupled in the Bellman optimality equation and we need to solve the equation for each state. In our problem, however, it turns out that we can obtain a closed-form expression of $w_i(\tau^{(i)})$ which tremendously reduces computation overhead. Before we proceed to the computation, we clarify that our problem indeed meets the assumption made by Whittle.

The applicability of the Whittle's index policy requires that each decoupled MDP in~\eqref{eq:decoupled mdp} is indexable. Denote $\mathbb{U}_i(w):=\{t:\pi^\star_i(t)=1,w_i=w\}$ as the set of states where transmission is optimal when the extra penalty is $w$.
\begin{definition}
A decoupled MDP is indexable if $\mathbb{U}_i(w)$ monotonically decreases from the whole state space $\mathbb{S}_i$ to the empty set as the extra cost $w_i$ increases from $-\infty$ to $+\infty$.
\end{definition}

The sensor scheduling problem is indeed indexable, which is based on the optimality of threshold policies and monotonicity of the threshold with respect to $w_i$.
\begin{lemma}\label{lemma: indexability}
\begin{enumerate}
	\item There exists a constant $\theta_i^\star(w_i)$ depending on $w_i$ such that the threshold policy of the form
	\begin{align*}
	\pi^\star_i(t) = \begin{cases} 1,&~\text{if}~t\geq\theta_i^\star(w_i),\\0,&~\text{if}~t\leq\theta_i^\star(w_i).\end{cases}
	\end{align*}
	achieves the minimization in~\eqref{eq:acoe for each} with $w=w_i$.
	\item The thresholds satisfy $\theta_i^\star(w_i)\leq\theta_i^\star(w_i')$ if $w_i\leq w'_i$.
\end{enumerate}
\end{lemma}
We conclude from Lemma~\ref{lemma: indexability} that the indexable condition indeed holds. As a threshold policy is optimal, we can obtain $U_i(w_i)=\{t:t \geq \theta_i^\star(w_i)\}$. From the monotonicity of the threshold, we can further obtain $U_i(w_i)\subset U_i(w'_i)$ if $w_i \geq w'_i$. Moreover, since $w_i=-\infty$ and $w_i=+\infty$ lead to $U_i(w_i)=\mathbb{S}_i$ and $U_i(w_i)=\emptyset$, we verify that the decoupled MDP for sensor $i$ is indexable.

Before we proceed to the closed-form expression for the Whittle's index, we need the following lemma to compute the averaged estimation error under a threshold policy.
\begin{lemma}\label{lemma: closed form expression for average costs}
The time-averaged communication rate under a threshold policy with threshold $\tau^{(i)}$ is
\begin{align*}
\lim_{T\to\infty} \frac{1}{T+1} \mathbb{E}\Big[\sum_{k=0}^T a_k^{(i)}\Big] = \frac{1}{\lambda_i\tau^{(i)}+1}.
\end{align*}
The time-averaged estimation error $J_e^{(i)}(\tau^{(i)})$ under the same threshold policy is
\begin{align*}
&J_e^{(i)}(\tau^{(i)}) = \\
&\begin{cases}
\lambda_i \Tr(S_{\overline{P}^{(i)}}) + (1-\lambda_i) \Tr(S_{Q_i}), &~\text{if}~\tau^{(i)}=0,\\
\Big[ \Tr(S_{h_i^{\tau^{(i)}}(\overline{P}^{(i)})}) + \frac{1-\lambda_i}{\lambda_i} \Tr(S_{Q_i}) \\ \quad  + \sum_{t=0}^{\tau^{(i)}-1}c_e^{(i)}(t)\Big] \cdot \frac{\lambda_i}{\lambda_i\tau^{(i)}+1},&~\text{if}~\tau^{(i)}>0,
\end{cases}
\end{align*}
where $S_{\overline{P}^{(i)}}$ and $S_{Q_i}$ are the solutions of
\begin{align*}
S = (1-\lambda_i)A_i S A_i^\top + \overline{P}^{(i)}
\end{align*}
and
\begin{align*}
S = (1-\lambda_i)A_i S A_i^\top + Q_i,
\end{align*}
respectively.
\end{lemma}
This lemma implies that, under a threshold policy, the time-averaged communication rate and the estimation error can be efficiently computed for each sensor $i$. This helps us develop an analytic expression of the Whittle's indices in the following.

\begin{theorem}\label{theorem: whittle's index}
The Whittle's index as a function of the time elapsed since the last successful transmission from sensor $i$ is
\begin{multline}
w_i(\tau^{(i)})=\frac{\lambda_i(\lambda_i\tau^{(i)}+1)}{1-\lambda_i} \\ \cdot\Big[ (\tau^{(i)}+1)J_e^{(i)}(\tau^{(i)}) - \sum_{t=0}^{\tau^{(i)}}c_e^{(i)}(t) \Big] - c_c^{(i)},\label{eq:form of whittle index}
\end{multline}
where $J_e^{(i)}(\tau^{(i)})$ is the expected time-averaged estimation error of sensor $i$ under a threshold policy with threshold $\tau^{(i)}$.
\end{theorem}

Theorem~\ref{theorem: whittle's index} gives an analytic expression of Whittle indices. Significant computation overhead is thus reduced compared with numerical algorithms such as value iteration and policy iteration. Moreover, this facilitates online implementations. It is worth noting that, apart from the extra penalty determined by the Whittle's index, every transmission will cause an energy cost $c_c^{(i)}$. Therefore, the Whittle's index can be negative. We revise the Whittle's index policy as follows. At each time step, we first pick $m$ sensors whose Whittle's indices are the top $m$, and then only schedule those sensors with positive Whittle's indices.

Weber and Weiss~\cite{weber1990index} proved that, if some conditions hold\footnote{The asymptotic optimality holds if the fluid approximation to the index policy has a globally asymptotically stable equilibrium point. The authors claims that examples violating these conditions are extremely rare and the suboptimality is expected to be minuscule.}, the Whittle's index policy is asymptotically optimal. The cost of the original MDP is lower bounded by the minimal average cost under a time-averaged constraint on its actions. As shown in~\eqref{eq: relaxed constraint}, the time-averaged constrained MDP is a relaxation of the original MDP, in which only $m$ out of $n$ sensors are scheduled at each time step. Meanwhile, as the Whittle's index policy meets the original constraint, it yields a performance upper bound of the original MDP. These bounds can be written as $C^{relax} \leq C^\star \leq C^{W}$, where $C^{relax}$ stands for the minimal cost under the relaxed MDP, $C^\star$ stands for the minimal cost for the original MDP, and $C^W$ stands for the time-averaged cost under the Whittle's index policy. Webber and Weiss showed that $C^W$ is asymptotically the same as $C^{relax}$ as $m$ and $n$ go to infinity with ratio $m/n$ fixed. Because $C^W$ asymptotically reaches $C^{relax}$, it also asymptotically reaches $C^\star$. In our numerical examples, the performance of the Whittle's index policy outperforms other two celebrated heuristics.
\section{NUMERICAL EXAMPLE}
In this section, we present numerical examples to illustrate the theoretical results. The first example is provided to show the optimality of monotone policies (Theorem~\ref{theorem: monotonicity for multiple processes}). The second example is provided to show the performance of the Whittle's index policy.

\begin{figure}[t]
	\subfigure[No transimission costs.]{
		\centering
		\includegraphics[width=0.22\textwidth]{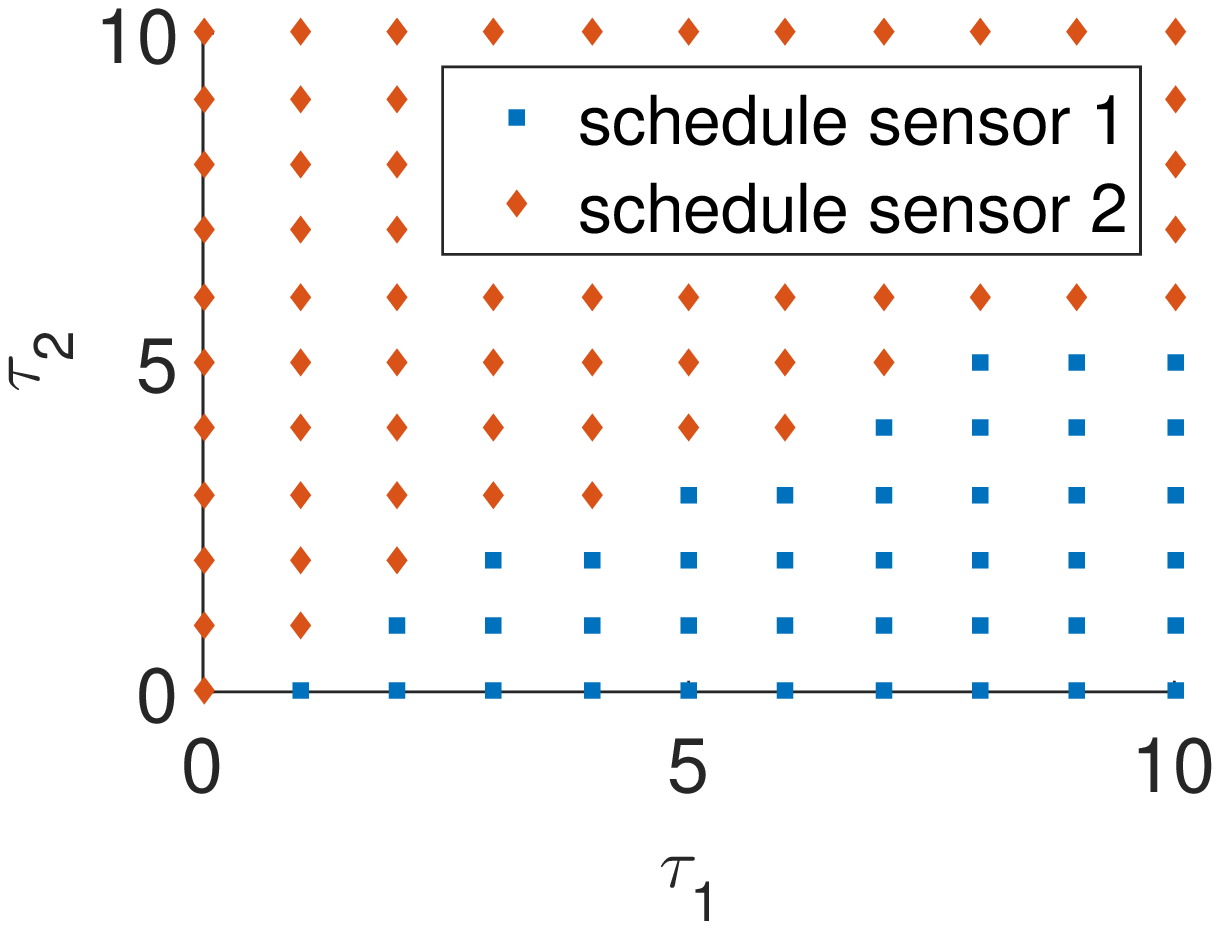}
	}
	\hspace{0.01em}
	\subfigure[With transmission costs.]{
		\centering
		\includegraphics[width=0.22\textwidth]{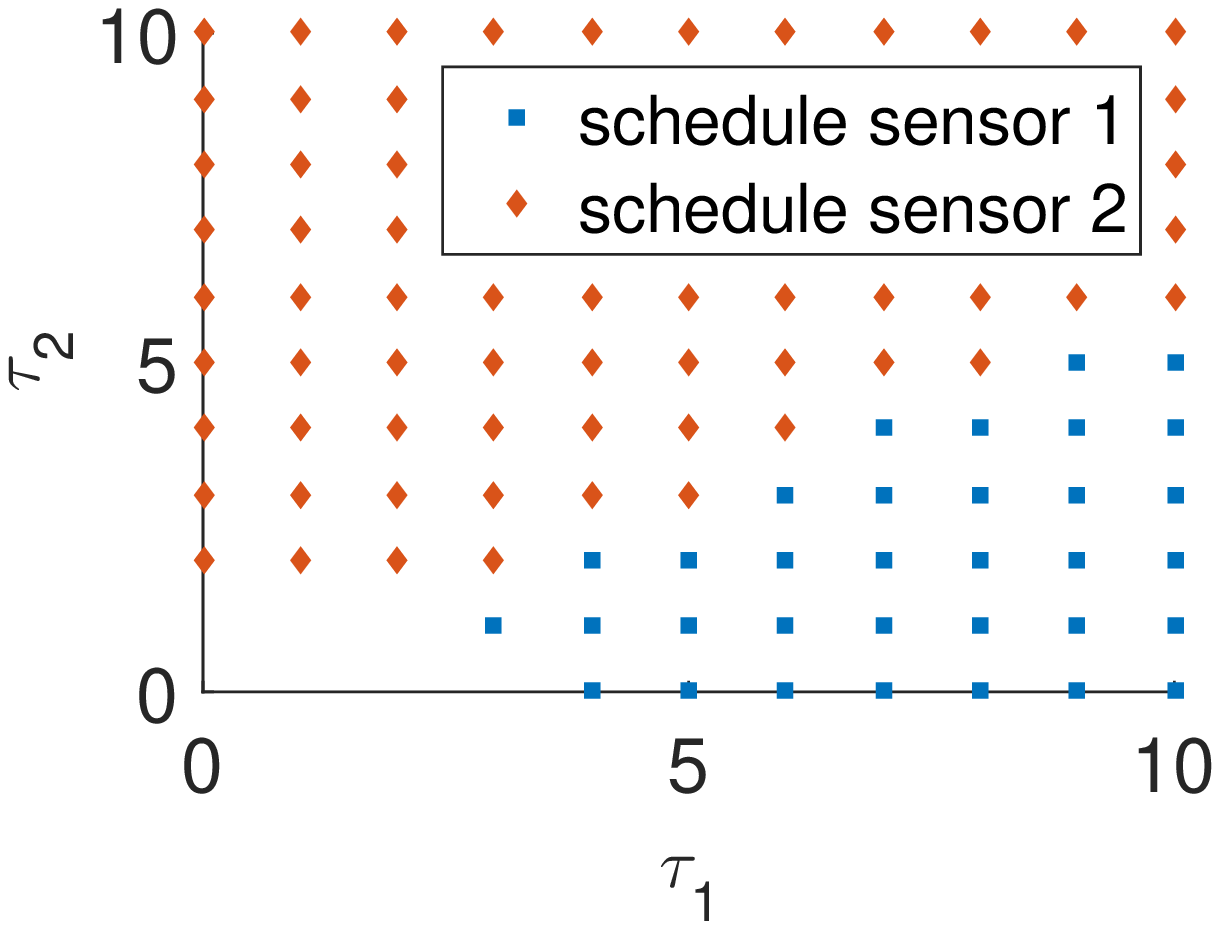}
	}
	\caption{Visualization of the monotone policy when $n=2$ and $m=1$.}
	\label{fig:threshold_cost}
\end{figure}

We first consider the case when $n=2$. The two processes and their parameters are as follows:
\begin{gather*}
{A}_1=\begin{bmatrix}1.1 &1\\0 &1\end{bmatrix}, ~{C}_1=\begin{bmatrix}2 &0\\0 &1\end{bmatrix}, ~{Q}_1=\begin{bmatrix}1 &0\\0 &1\end{bmatrix}, ~{R}_1=\begin{bmatrix}1 &0\\0 &1\end{bmatrix};\\
{A}_2=\begin{bmatrix}1 &1\\0 &1.2\end{bmatrix}, ~{C}_2=\begin{bmatrix}1 &0\\0 &1\end{bmatrix}, ~{Q}_2=\begin{bmatrix}1 &0\\0 &1\end{bmatrix}, ~{R}_2=\begin{bmatrix}1 &0\\0 &1\end{bmatrix}.
\end{gather*}
Moreover, the packet arrival rate of the two channels are $\lambda_1=0.8$ and $\lambda_2=0.9$, respectively. We consider two scenarios with zero or positive transmission costs, respectively. For the positive costs, we let $c_c^{(1)}=20$ and $c_c^{(2)}=10$. We use the relative value iteration to compute an optimal policy. The monotonicity structure of the optimal policy is shown in Fig.~\ref{fig:threshold_cost}. Sub-figure (a) shows an optimal policy when $c_c^{(1)}= c_c^{(2)}=0$, and Sub-figure (b) shows an optimal policy when $c_c^{(1)}=20$ and $c_c^{(2)}=10$. The horizontal and vertical axes represent the consecutive packet drops of sensor $1$ and $2$, respectively. It is clear that there exists a boundary splitting the $(\tau_1,\tau_2)$ plane into two regions. The states in the left upper corner correspond to scheduling sensor $2$, while the states in the right lower corner correspond to scheduling sensor $1$. In addition, when there are extra transmission costs, it may be optimal not to schedule any sensor if $\tau^{(i)}$ are small.

\begin{figure*}[t]
	\includegraphics[width=0.9\textwidth]{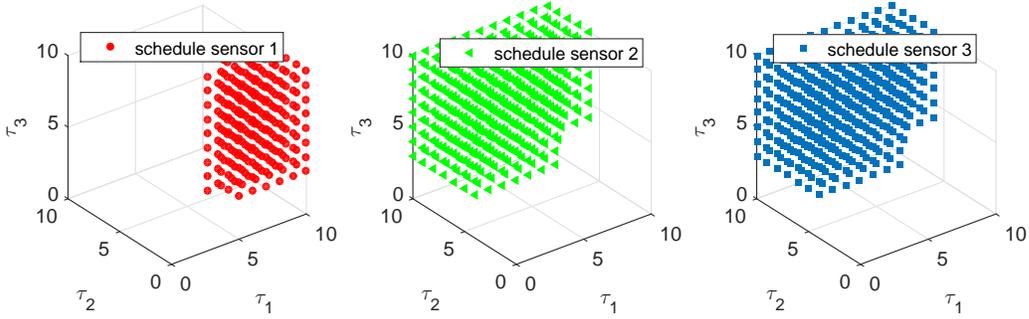}
	\caption{Visualization of the switching surface policy when $n=3$, $m=2$ and communication costs $c_c^{(1)}=50$, $c_c^{(2)}=30$, $c_c^{(3)}=40$.}
	\label{fig:swithcingsurface_cost}
\end{figure*}

When $n>2$, the monotone structure is hard to depict. We consider a case with $n=3$ and $m=2$. The LTI processes dynamics are as follows:
\begin{gather*}
{A}_1=\begin{bmatrix}1.1 &1\\0 &1\end{bmatrix}, ~{C}_1=\begin{bmatrix}1 &0\end{bmatrix}, ~{Q}_1=\begin{bmatrix}1 &0\\0 &4\end{bmatrix}, ~{R}_1=1;\\
{A}_2=\begin{bmatrix}1.2 &1\\0 &1\end{bmatrix}, ~{C}_2=\begin{bmatrix}1 &0\end{bmatrix}, ~{Q}_2=\begin{bmatrix}1 &0\\0 &2\end{bmatrix}, ~{R}_2=1;\\
{A}_3=\begin{bmatrix}1.1 &1\\0 &1.3\end{bmatrix}, ~{C}_3=\begin{bmatrix}1 &0\\0 &1\end{bmatrix}, ~{Q}_3=\begin{bmatrix}1 &0\\0 &1\end{bmatrix}, ~{R}_2=I,
\end{gather*}
where $I = \begin{bmatrix}1 &0\\0 &1\end{bmatrix}$. The packet arrivals are set as $\lambda_i=0.9$ for $i=1,2,3$. Let the communication costs be $c_c^{(1)}=50$, $c_c^{(2)}=30$, $c_c^{(3)}=40$. There are seven feasible actions.
\begin{enumerate}
	\item No schedule for any sensor;
	\item Schedule one sensor: schedule sensor 1, schedule sensor 2, schedule sensor 3;
	\item Schedule two sensors: schedule sensor 1 and 2, schedule sensor 1 and 3, schedule sensor 2 and 3.	
\end{enumerate}
By following the same procedure when $n=2$, we obtain an optimal policy. For each sensor, either it is scheduled or not is a feasible action.  We plot optimal actions for each sensor with respect to different states in Fig.~\ref{fig:swithcingsurface_cost}. The region of scheduling each sensor are shown in each sub-figure. We can observe that there exists a switching surface between scheduling a particular sensor and not scheduling this sensor. As there are extra communication costs, we can see that it is optimal to schedule no sensors when $\tau^{(i)}$ are small.

\begin{figure}[t]
	\centering
	\includegraphics[width=0.4\textwidth]{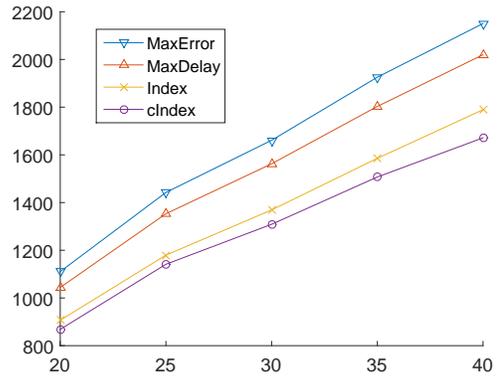}
	\caption{Performance comparison of heuristic policies.}
	\label{fig:performance comparison}
\end{figure}

\begin{figure}[t]
	\centering
	\includegraphics[width=0.4\textwidth]{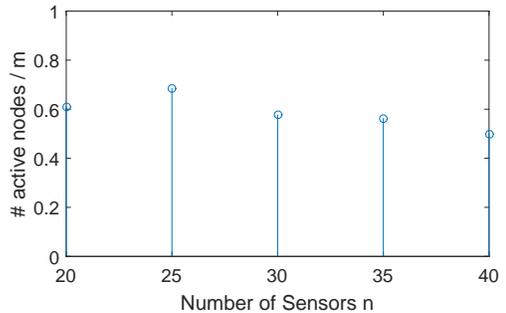}
	\caption{The ratio between the average number of active sensors over the allowed simultaneous transmissions $m$.}
	\label{fig:activepercent}
\end{figure}

Finally, we present the performance of Whittle's index policy. For comparison, we also simulate scheduling under two celebrated heuristics, \emph{maximum-error-first} policy and \emph{maximum-delay} first policy. In the former, we choose the $m$ sensors whose expected errors $\Tr(h_i^{(\tau^{(i)}_k)}(\overline{P}^{(i)}))$ are the $m$-largest at time $k$. In the later, we choose the $m$ sensors whose delays $\tau^{(i)}_k$ are the $m$-largest. Since there are transmission costs, the Whittle's index may not be positive. We consider two types of Whittle's index policy, the original one and the revised one we discussed in the end of the last section. We randomly generate $40$ first-order LTI systems:
\begin{align*}
x^{(i)}_{k+1} = Ax^{(i)}_{k} + w^{(i)}_{k}, y^{(i)}_{k} = Cx^{(i)}_{k} + v^{(i)}_{k},
\end{align*}
with system gains $A$ drawn from a standard normal distribution, observation gains $C$ drawn from uniform distribution on the closed interval $[1,10]$, and the state disturbance covariances $\mathbb{E}[w_k^{(i)}\cdot w_k^{(i)}]$ and the observation disturbance covariances $\mathbb{E}[v_k^{(i)}\cdot v_k^{(i)}]$ drawn from uniform distribution on the closed interval $[0, 100]$. The transmission costs are randomly drawn from the closed interval $[5,15]$. We simulate five scenarios, $n=20$ with $m=8$, $n=25$ with $m=10$, $n=30$ with $m=12$, $n=35$ with $m=14$, and $n=40$ with $m=16$. The ratio $\frac{m}{n}=0.4$ in all scenarios. In each scenario, we run Monte Carlo simulations of the scheduling process of the four scheduling heuristics over a time-horizon with length $1000$ for $100$ times. We compute the averaged total costs of each heuristics, which consist of the averaged estimation error and the averaged transmission costs. The performance of each heuristics is shown in Fig.~\ref{fig:performance comparison}, where ``MaxError" refers to the maximum-error-first policy, `MaxDelay" refers to the maximum-delay-first policy, and ``Index" and ``cIndex" refers to the original Whittle's index policy and revised Whittles' index policy, respectively. We observe that the two Whittle's index policies outperform the other two heuristics. The revised policy in most cases performs better than the original one as the costs of transmission are also considered. The average percentage of active sensor nodes under the revised policy is reported in Fig.~\ref{fig:activepercent}. Note that the percentage of other three policies is always one as they always schedule $m$ sensors simultaneously.

\section{CONCLUSION}
We formulated the multiple sensor scheduling problem as a Markov decision process (MDP) with an average cost over an infinite horizon. An algorithm (Algorithm~\ref{alg:feasibility}) was proposed to check the existence of a deterministic stationary optimal policy. We proved the optimality of monotone policies. The monotone structure reduced the computation effort of finding an optimal policy and facilitated online implementation. We leveraged the structure of the problem to prove that each process is indexable in the sense of Whittle's. We adopted Whittle's index to construct an index heuristics with closed-form expressions, which tremendously saved computation effort and facilitated online implementation. Numerical examples showed the empirical performance of the proposed index policy outperforms other two common heuristics.

The current setup assumes that the channel condition is invariant and known beforehand. It would be a challenging problem if the channel condition follows a time-varying model and the parameters are unknown. In this case, a learning based method such as $Q$-learning can be used. In this work, the centralized scheduling is considered. Another future direction involves a distributed design. If some information exchange among the sensors is applicable, the scheduling policy can be done in a distributed manner.

\bibliographystyle{IEEETran}
\bibliography{IEEEabrv,blreference}

\section*{Appendix}

\subsection{Proof of Theorem~\ref{thm:existence}}
Based on~\cite[Theorem 5.5.4]{hernandez1995discrete}, if there exists a policy $\{\pi_k\}_{k=0}^\infty\in\Pi$ such that
\begin{align}\label{eq: ds condition}
\lim_{T\to \infty} \frac{1}{T+1} \sum_{k=0}^{T} \sum_{i=1}^{n} \mathbb{E} [c_e^{(i)}(\tau_k^{(i)})+  c_c^{(i)}a_k^{(i)}] < \infty,
\end{align}
i.e., the corresponding average cost is bounded, then the optimal scheduling policy can be found in the set of deterministic stationary policies.
\begin{lemma}\label{lemma: existence of acoe}
	If there exists a policy $\{\pi_k\}_{k=0}^\infty\in\Pi$ such that~\eqref{eq: ds condition} holds, then there exist a constant $\mathcal{J}^\star$, a function $V^\star(\tau)$, and a deterministic stationary policy $\pi^\star:\mathbb{S}\mapsto\mathbb{A}$ that satisfy the following Bellman optimality equation
	\begin{align}\label{eq:acoe}
	\mathcal{J}^\star + V^\star(s) = \min_{a\in\mathbb{A}} \Bigg[ c(s,a) + \sum_{s_+\in\mathbb{S}} V^\star(s_+)\mathtt{Pr}(s_+|s,a) \Bigg]
	\end{align}
	and
	\begin{align*}
	\mathcal{J}^\star + V^\star(s) = \Bigg[ c(s,\pi^\star(s)) + \sum_{\tau_+\in\mathbb{S}} V^\star(s_+)\mathtt{Pr}(s_+|s,\pi^\star(s)) \Bigg].
	\end{align*}
	In addition,
	\begin{align*}
	J(\pi^\star) = \min_{\pi\in\Pi} J(\pi) = \mathcal{J}^\star.
	\end{align*}
\end{lemma}

\begin{IEEEproof}
The proof relies on the vanishing discount approach in~\cite[Theorem 5.5.4]{hernandez1995discrete}. We define the discounted total cost
\begin{align*}
V^\star_{\beta}(s) = \min_{\pi\in\Pi} \lim_{T\to\infty} \mathbb{E} \Big[ \sum_{k=0}^T  \beta^kc(s_k,a_k) \mid s_0 = s \Big],
\end{align*}
for an auxiliary purpose. In summary, the following conditions need to be verified.
\begin{enumerate}
\item The one-stage cost $c(s,a)$ is continuous, nonnegative, and for any $r\in\mathbb{R}$ the set $\{a\in\mathbb{A}| c(s,a)<r\}$ is compact.
\item The probability transition law $\mathtt{Pr}(s_+|s,a)$ is strongly continuous.
\item There exists a state $z\in\mathbb{S}$, a number $0<\underline{\beta}<1$ and $\overline{M}\geq 0$ such that
\begin{align*}
(1-\beta) V^\star_{\beta}(z) \leq \overline{M},~\forall s\in\mathbb{S},~ \underline{\beta}\leq\beta<1.
\end{align*}
\item There exists a constant $\underline{M}\geq 0$ and a nonnegative function $b(s)$ on $\mathbb{S}$ such that
\begin{align*}
-\underline{M} \leq V^\star_{\beta}(s)-V^\star_{\beta}(z) \leq b(s),~\forall s\in\mathbb{S},~ \underline{\beta}\leq\beta<1.
\end{align*}
\item The function $b(s)$ above is measurable and for any $s\in\mathbb{S}$ and $a\in\mathbb{A}$: $\sum_{s_+\mathbb{S}} b(s_+)\mathbb{P}(s_+ |s,a)<\infty$
\item The sequence $\{ V^\star_{\beta(n)}(s)-V^\star_{\beta(n)}(z)\}$ is equicontinuous.
\end{enumerate}
The first two conditions are satisfied in this problem as the action space consists of finite actions and the one-stage cost is bounded below by zero. If there exists a $\pi\in\Pi$ policy such that the average cost is bounded, i.e.,
\begin{align*}
\lim_{T\to\infty} \frac{1}{T+1} \sum_{k=0}^T \mathbb{E} [c(s_k,\pi(s_k))] < \infty.
\end{align*}
By Abelian Theorem~\cite[Lemma 5.3.1]{hernandez1995discrete}, we have
\begin{align*}
&\liminf_{T\to\infty} \frac{1}{T+1} \sum_{k=0}^T  \mathbb{E} [c(s_k,\pi(s_k))]\\
\leq &\liminf_{\beta\to 1} (1-\beta)\sum_{k=0}^\infty \beta^k\mathbb{E}[c(s_k,\pi(s_k))]\\
\leq &\limsup_{\beta\to 1} (1-\beta)\sum_{k=0}^\infty  \beta^k\mathbb{E}[c(s_k,\pi(s_k))]\\
\leq &\limsup_{T\to\infty} \frac{1}{T+1} \sum_{k=0}^T \mathbb{E} [c(s_k,\pi(s_k))].
\end{align*}
As the limit $$ \lim_{T\to\infty} \frac{1}{T+1} \sum_{k=0}^T \mathbb{E} [c(s_k,\pi(s_k))]$$ exists, the limit $$\lim_{\beta\to 1} (1-\beta)\sum_{k=0}^\infty \beta^k\mathbb{E}[c(\tau_k,f(\tau_k))]$$ also exists. Denote this limit as $M_c$. The existence of the limit implies that for every $\underline{\beta}<1$, there exists $\varepsilon \geq 0$ such that $| (1-\beta)\sum_{k=0}^\infty \beta^k\mathbb{E}[c(s_k,\pi(s_k))] - M_c | \leq \varepsilon$. Therefore, we can derive
\begin{align*}
(1-\beta)V^\star_{\beta}(s) &\leq (1-\beta)\sum_{k=0}^\infty \beta^k\mathbb{E}[c(s_k,\pi(s_k)) \mid s_0=s] \\ &\leq M_c+\varepsilon
\end{align*}
for any $s\in\mathbb{S}$ and $\underline{\beta}  \leq \beta < 1$, which verifies condition (3). By~\cite[Lemma 7.4.1]{sennott2009stochastic}, condition (4) also holds.
Since there are finite possible $s_+$ in $\mathtt{Pr}(s_+|s,a)$ given $s\in\mathbb{S}$, condition (5) also holds. Lastly, as the state space is discrete, condition (6) also holds. As the six conditions are satisfied, the result holds.
\end{IEEEproof}

Lemma~\ref{lemma: existence of acoe} serves as a guide for us to establish a sufficient condition of existence of a regular optimal policy. In brevity, we want to find a stationary and deterministic policy such that the associated time average cost is bounded. This can be done using the results from~\cite[Theorem 3]{mesquita2012redundant}.

The setup in~\cite{mesquita2012redundant}, however, is different from the setting in this work. They assume that the sensors may send redundant local estimate through multiple channels simultaneously and thus their approach is not directly applicable in this work. If the allowable channel number is one, i.e., $m=1$, we can immediately obtain that if 
\begin{align}\label{eq:one channel stability condition}
\max_{i}\rho^2(A_i)\max_j(1-\lambda_j)<1
\end{align}
for $1 \leq i\leq n$ and $1\leq j \leq n$, the time average of the sum of the estimation error covariance of all sensors is bounded under an $L$-triggered policy\footnote{The term $L$-triggered policy comes from~\cite{mesquita2012redundant}. In this work, it only schedules sensors with $\tau^{(i)} > L$.}.

We generalize this result for $m>1$. The idea is as follows. We can partition the $n$ processes into $m$ groups. At each time step, only one sensor in each group is allowed to transmit packets. Then the boundedness condition turns out to be whether there exists a partition such that the time-averaged cost of each group are bounded. Note that the partition is applied to the unstable processes because the boundedness holds even if the stable processes are never scheduled.

If the output of Algorithm~\ref{assumption:neccessary for stability} is less than $m$, we can partition the $n$ processes into $m$ groups, i.e., $\{\mathcal{N}_1,\dots,\mathcal{N}_m\}$. In each group $\mathcal{N}_j$, there exists an $L_j$ such that an $L_j$-triggered policy leads to a bounded average estimation error. In addition, time-averaged communication costs are always bounded. Therefore, there exists a policy such that~\eqref{eq: ds condition} holds, which shows the optimality of a deterministic stationary policy.

\subsection{Proof of Theorem~\ref{theorem: monotonicity for multiple processes}}
Before we proceed to the proof, we make two definitions. For every process $i$, we can define a partial order $\leq_i$ on the states $s$. The same convention of partial order is defined for the actions $a$.

The monotonicity for every process $i$ can be perceived as monotonicity of the optimal action on the state space. This can be guaranteed if the following four conditions hold.
\begin{enumerate}
	\item If $s\leq_i s'$, $c(s,a)\leq c(s',a)$ for any $a\in\mathbb{A}$;
	\item If $s \leq_i s'$, for any $a\in\mathbb{A}$,
	\begin{align*}
	\sum_{s_+}\mathtt{Pr}(s_+|s,a)V(s_+) \leq \sum_{s_+}\mathtt{Pr}(s_+|s',a)V(s_+)
	\end{align*}
	where $V(s)$ is any monotone increasing function, i.e., $V(s)\leq V(s')$ if $s\leq_is'$;
	\item If $s \leq_i s'$ and $a \leq_i a'$, $c(s,a) + c(s',a') \leq c(s',a) + c(s,a')$;
	\item If $s \leq_i s'$ and $a \leq_i a'$,
	\begin{align*}
	&\sum_{s_+}\mathtt{Pr}(s_+|s,a)V(s_+) + \sum_{s_+}\mathtt{Pr}(s_+|s',a')V(s_+)\\ \leq &\sum_{s_+}\mathtt{Pr}(s_+|s',a)V(s_+) + \sum_{s_+}\mathtt{Pr}(s_+|s,a')V(s_+),
	\end{align*}
	where $V(s)$ is again any monotonically increasing function.
\end{enumerate}

Conditions (1) and (2) address the monotonicity of $c(s,a)$ and $\mathtt{Pr}(s_+|s,a)$, while conditions (3) and (4) the submodularity of $c(s,a)$ and $\mathtt{Pr}(s_+|s,a)$. Consider a discounted cost MDP over a finite time-horizon
\begin{align*}
\min_{\pi\in\Pi}  \mathbb{E} \Big[\sum_{k=0}^{T}  \beta^kc(s_k,a_k) \Big].
\end{align*}
An optimal policy must satisfy the following Bellman optimality equation defined backwards (from $k=T$ to $k=0$) by
\begin{align*}
V_{T,\beta}^\star(s) := \min_{a\in\mathbb{A}} c(s,a)
\end{align*}
and for $t=T-1,T-2,\dots,0$,
\begin{align*}
V_{k,\beta}^\star(s) := \min_{a\in\mathbb{A}} \Big[c(s,a) +  \beta \sum_{s_+}\mathtt{Pr}(s_+|s,a)V_{k+1,\beta}^\star(s_+) \Big].
\end{align*}
If the above four conditions are satisfied, the quantity inside the minimization of the Bellman optimality equation $c(s,a)+\beta \sum_{s_+}\mathtt{Pr}(s_+|s,a)V(s_+)$ is monotone and submodular in $s$ and $a$, which shows that there exists a monotone policy being an optimal policy for any finite-horizon MDP. By again using the vanishing discount approach~\cite[Theorem 5.5.4]{hernandez1995discrete}, the monotonicity is propagated to the time-averaged MDP. The proof of Lemma~\ref{lemma: existence of acoe} has already verified the applicability of such an argument. The remaining task is to verify the four conditions.

Conditions (1) and (3) are satisfied according to the definition of $c(s,a)$. Denote $\tau^{(-i)}=(\tau^{(j)})_{j\neq i}$ and $a^{(-i)}=(a^{(j)})_{j\neq i}$ as the states and actions of all sensors except sensor $i$. Note that
\begin{align*}
&\sum_{s_+}\mathtt{Pr}(s_+|s,a)V(s_+)\\
= &\sum_{\tau^{(i)}_+} \mathtt{Pr}^{(i)}(\tau^{(i)}_+|\tau^{(i)},a^{(i)})\sum_{\tau^{(-i)}_+} \mathtt{Pr}^{(-i)}(\tau^{(-i)}_+|\tau^{(-i)},a^{(-i)})V(s_+)\\
=&\sum_{\tau^{(i)}_+} \mathtt{Pr}^{(i)}(\tau^{(i)}_+|\tau^{(i)},a^{(i)})\tilde{V}(\tau^{(i)}_+),
\end{align*}
where $\tilde{V}(\tau^{(i)}_+):=\sum_{\tau^{(-i)}_+} \mathtt{Pr}^{(-i)}(\tau^{(-i)}_+|\tau^{(-i)},a^{(-i)})V(s_+)$ is monotone in $\tau^{(i)}_+$. By its definition in~\eqref{eq:single process transisiton}, the transition probability $\mathtt{Pr}^{(i)}(\tau^{(i)}_+|\tau^{(i)},a^{(i)})$ satisfies
\begin{align*}
\sum_{\tau^{(i)}_+}\mathtt{Pr}(\tau^{(i)}_+|\tau^{(i)},a^{(i)})\tilde{V}(\tau^{(i)}_+) \leq \sum_{\tau^{(i)}_+}\mathtt{Pr}(\tau^{(i)}_+|\tau'^{(i)},a^{(i)})\tilde{V}(\tau^{(i)}_+)
\end{align*}
for any $a^{(i)}\in\mathbb{A}_i$ if $\tau^{(i)}\leq \tau'^{(i)}$; and
\begin{align*}
&\sum_{\tau^{(i)}_+}\mathtt{Pr}(\tau^{(i)}_+|\tau^{(i)},a^{(i)})\tilde{V}(\tau^{(i)}_+) + \sum_{\tau^{(i)}_+}\mathtt{Pr}(\tau^{(i)}_+|\tau'^{(i)},a'^{(i)})\tilde{V}(\tau^{(i)}_+) \\&\leq \sum_{\tau^{(i)}_+}\mathtt{Pr}(\tau^{(i)}_+|\tau'^{(i)},a^{(i)})\tilde{V}(\tau^{(i)}_+) + \sum_{\tau^{(i)}_+}\mathtt{Pr}(\tau^{(i)}_+|\tau^{(i)},a'^{(i)})\tilde{V}(\tau^{(i)}_+)
\end{align*}
if $\tau^{(i)} \leq \tau'^{(i)}$ and $a^{(i)} \leq a'^{(i)}$. This shows that conditions (2) and (4) are also satisfied, which completes the proof.

\subsection{Proof of Lemma~\ref{lemma: indexability}}
\emph{Part I. Optimality of threshold policy.} The threshold policy can be perceived as a monotone policy, whose optimality can be verified by the following four conditions as follows, which are similar to that mentioned in the proof of Theorem~\ref{theorem: monotonicity for multiple processes}.
\begin{enumerate}
	\item If $\tau^{(i)}\leq \tau'^{(i)}$, $c^{(i)}(\tau^{(i)},a^{(i)}) \leq c^{(i)}(\tau'^{(i)},a^{(i)})$ for any $a^{(i)}\in\mathbb{A}_i$;
	\item If $\tau^{(i)}\leq \tau'^{(i)}$, 
	\begin{align*}
	\sum_{\tau^{(i)}_+ \geq t}\mathtt{Pr}(\tau^{(i)}_+|\tau^{(i)},a^{(i)}) \leq \sum_{\tau^{(i)}_+\geq t}\mathtt{Pr}(\tau^{(i)}_+|\tau'^{(i)},a^{(i)})
	\end{align*}
	for any $a^{(i)}\in\mathbb{A}_i$ and $t\in\mathbb{S}_i$;
	\item If $\tau^{(i)} \leq \tau'^{(i)}$ and $a^{(i)} \leq a'^{(i)}$, $c^{(i)}(\tau^{(i)},a^{(i)}) + c^{(i)}(\tau'^{(i)},a'^{(i)}) \leq c^{(i)}(\tau'^{(i)},a^{(i)}) + c^{(i)}(\tau^{(i)},a'^{(i)})$;
	\item If $\tau^{(i)} \leq \tau'^{(i)}$ and $a^{(i)} \leq a'^{(i)}$,
	\begin{align*}
	&\sum_{\tau^{(i)}_+ \geq t}\mathtt{Pr}(\tau^{(i)}_+|\tau^{(i)},a^{(i)}) + \sum_{\tau^{(i)}_+ \geq t}\mathtt{Pr}(\tau^{(i)}_+|\tau'^{(i)},a'^{(i)}) \\&\leq \sum_{\tau^{(i)}_+\geq t}\mathtt{Pr}(\tau^{(i)}_+|\tau'^{(i)},a^{(i)}) + \sum_{\tau^{(i)}_+\geq t}\mathtt{Pr}(\tau^{(i)}_+|\tau^{(i)},a'^{(i)})
	\end{align*}
	for all $t$;
\end{enumerate}
where $c^{(i)}(\tau^{(i)},a^{(i)})=c_e^{(i)}(\tau^{(i)}) + c_c^{(i)}a^{(i)} + w_i a^{(i)}$. Conditions (1) and (3) can be seen from the definition of $c^{(i)}(\tau^{(i)},a^{(i)})$ and conditions (2) and (4) can be verified through calculation. Since the four conditions are satisfied, the optimality of monotone policy holds for any finite-horizon MDPs. By using the vanishing discount argument, the monotone policy is preserved for the time-averaged MDP.

\emph{Part II. Monotonicity of the optimal threshold.} We need the following lemma to prove the monotonicity of the optimal threshold.
\begin{lemma}\label{lemma:argmin preserves submodularity}
Let $f:\mathbb{X}\times\mathbb{Y}\to\mathbb{R}$ be a submodular function, i.e.,
\begin{align*}
f(x^+,y^+) + f(x^-,y^-) \leq f(x^+,y^-) + f(x^-,y^+)
\end{align*}
if $x^+\geq x^-$ and $y^+\geq y^-$. The function 
\begin{align*}
g(x) := \max\{y^\star\in\argmin_{y\in\mathbb{Y}} f(x,y)\}
\end{align*}
is increasing in $x$. 
\end{lemma}
\begin{IEEEproof}
Suppose $x^+\geq x^-$. As $f(x,y)$ is submodular, for any $y\leq g(x^-)$, we have
	\begin{align*}
	f(x^+, g(x^-)) - f(x^+, y) \leq f(x^-, g(x^-)) - f(x^-, y) \leq 0,
	\end{align*}
	which implies $f(x^+, g(x^-)) \leq f(x^+, y)$
	for any $y\leq g(x^-)$. Therefore, $g(x^+) \geq g(x^-)$.
\end{IEEEproof}

Consider the total time-averaged cost 
\begin{align*}
\mathcal{J}_i(w_i,\theta_i) = \lim_{T\to\infty} \frac{1}{T+1}\mathbb{E}^{\theta_i}\Big[\sum_{k=0}^{T} c_e^{(i)}(\tau^{(i)}_k) +  c_c^{(i)}a^{(i)}_k + w_ia^{(i)}_k \Big],
\end{align*}
where $\mathbb{E}^{\theta_i}$ stands for the expectation under a threshold policy with threshold $\theta_i$. It suffices to prove that $\mathcal{J}_i(w_i,\theta_i)$ is submodular in $w_i$ and $\theta_i$. Given a threshold policy $\theta_i$, we can compute the stationary distribution of the states of arm $i$ as follows.
\begin{align}
\pi_i(\tau^{(i)};\theta_i)=
\begin{cases}
\frac{\lambda_i}{\lambda_i\theta_i+1},&\text{if~} \tau^{(i)}\leq\theta_i,\\
\frac{\lambda_i}{\lambda_i\theta_i+1}(1-\lambda_i)^{\tau^{(i)}-\theta_i},&\text{if~} \tau^{(i)}>\theta_i,\\
0,&\text{otherwise}.
\end{cases}\label{eq: ipm with threshold policy}
\end{align}
Therefore, we can obtain 
\begin{align}
\lim_{T\to\infty} \frac{1}{T+1}\mathbb{E}^{\theta_i}\Big[\sum_{k=0}^{T}w_ia^{(i)}_k\Big]&=w_i (1-\theta_i\frac{\lambda_i}{\lambda_i\theta_i+1})\nonumber\\
&=\frac{w_i}{\lambda_i\theta_i+1}.\label{eq: average penalty}
\end{align}
This quantity is submodular in $w_i$ and $\theta_i$ because, if $w_i\geq w'_i$ and $\theta_i\leq \theta'_i$, we can obtain
\begin{align*}
&\frac{w_i}{\lambda_i\theta_i+1} + \frac{w'_i}{\lambda_i\theta'_i+1} - \frac{w_i}{\lambda_i\theta'_i+1} - \frac{w'_i}{\lambda_i\theta_i+1}\\
=&\frac{(w-w')(\lambda_i\theta'_i-\lambda_i\theta_i)}{(\lambda_i\theta_i+1)(\lambda_i\theta'_i+1)}\geq 0,
\end{align*}
which is equivalent to
$$
\frac{w_i}{\lambda_i\theta_i+1} + \frac{w'_i}{\lambda_i\theta'_i+1} \geq \frac{w_i}{\lambda_i\theta'_i+1} + \frac{w'_i}{\lambda_i\theta_i+1}.
$$
As the quantity
\begin{align*}
\lim_{T\to\infty} \frac{1}{T+1}\mathbb{E}^{\theta_i}\Big[\sum_{k=0}^{T} c_e^{(i)}(\tau^{(i)}_k)\Big]
\end{align*}
only depends on $\theta_i$, the total averaged cost $\mathcal{J}(w_i,\theta_i)$ is submodular in $w_i$ and $\theta_i$. Therefore, by Lemma~\ref{lemma:argmin preserves submodularity}, $\theta_i^{\star}(w_i)$ monotonically increases with respect to $w_i$.

\subsection{Proof of Lemma~\ref{lemma: closed form expression for average costs}}
In~\eqref{eq: average penalty} in the proof of Lemma~\ref{lemma: indexability}, we can obtain the time-averaged communication rate under a threshold policy with threshold $\tau^{(i)}$ is
\begin{align*}
\lim_{T\to\infty} \frac{1}{T+1}\mathbb{E}\Big[\sum_{k=0}^{T}a^{(i)}_k\Big]=\frac{1}{\lambda_i\tau^{(i)}+1}.
\end{align*}

To compute the time-averaged estimation error $J_e^{(i)}(\tau^{(i)})$, we need the following lemma regarding computation of a Lyapunov equation.
\begin{lemma}\cite[Lemma D.1.2]{kailath2000linear}
For a given positive definite symmetric $X$, there exists a unique positive definite symmetric $S$ satisfying $S = ASA^\top + X$ if and only $\rho(A)<1$, where $\rho(A)$ is the spectral radius of $A$. In addition, the unique $S$ can be computed by
\begin{align*}
S = \sum_{t=0}^\infty A^tX(A^\top)^t.
\end{align*}
\end{lemma}
The time-averaged estimation error $J_e^{(i)}(\tau^{(i)})$ can be computed by
\begin{align*}
J_e^{(i)}(\tau^{(i)}) = \sum_{t=0}^\infty \pi_i(t;\tau^{(i)})c_e^{(i)}(t),
\end{align*}
where $\pi_i(t;\tau^{(i)})$ is defined in~\eqref{eq: ipm with threshold policy}.

When $\tau^{(i)}=0$, we have
\begin{align*}
&J_e^{(i)}(0)\\
= &\sum_{t=0}^\infty \lambda_i(1-\lambda_i)^t \Tr[h_i^t(\overline{P}^{(i)})]\\
= &\lambda_i \Tr\Big\{ \sum_{t=0}^\infty \Big[ (1-\lambda_i)^tA_i^{t}\overline{P}^{(i)}(A_i^\top)^t \\
&+ \sum_{k=0}^t (1-\lambda_i)^{k+1}A_i^kQ_i(A_i^\top)^k \Big] \Big\}\\
=& \lambda_i \Tr\Big\{ \sum_{t=0}^\infty  (1-\lambda_i)^tA_i^{t}\overline{P}^{(i)}(A_i^\top)^t \Big\} \\
&+ \lambda_i \Tr\Big\{ \sum_{t=1}^\infty(1-\lambda_i)^t\sum_{k=0}^\infty (1-\lambda_i)^kA_i^kQ_i(A_i^\top)^k
\Big\}\\
= &\lambda_i \Tr(S_{\overline{P}^{(i)}}) + \lambda_i\Tr\Big\{ \sum_{t=1}^\infty(1-\lambda_i)^tS_{Q_i}\Big\}\\
= &\lambda_i \Tr(S_{\overline{P}^{(i)}}) + (1-\lambda_i)\Tr(S_{Q_i}).
\end{align*}
When $\tau^{(i)}>0$, note that
\begin{align*}
&J_e^{(i)}(\tau^{(i)})\\
= &\sum_{t=0}^{\tau^{(i)}-1} \frac{\lambda_i}{\lambda_i\tau^{(i)}+1}\Tr[h_i^t(\overline{P}^{(i)})] \\&
+ \sum_{t=0}^\infty \frac{\lambda_i}{\lambda_i\tau^{(i)}+1}(1-\lambda_i)^t \Tr[h_i^t(h_i^{\tau^{(i)}}(\overline{P}^{(i)}))].
\end{align*}
Note that the form of the second infinite summation is the same as that in $J_e^{(i)}(0)$. We can therefore obtain
\begin{align*}
J_e^{(i)}(\tau^{(i)}) = \Big[ \Tr(S_{h_i^{\tau^{(i)}}(\overline{P}^{(i)})}) + \frac{1-\lambda_i}{\lambda_i} \Tr(S_{Q_i}) \\ \quad + \sum_{t=0}^{\tau^{(i)}-1}c_e^{(i)}(t)\Big] \cdot \frac{\lambda_i}{\lambda_i\tau^{(i)}+1}
\end{align*}
for $\tau^{(i)}>0$.

\subsection{Proof of Theorem~\ref{theorem: whittle's index}}
By its definition, the Whittle's index $w_i(\tau^{(i)})$ should be such that the expected costs of being passive (no transmission) and be active (transmission) are equal under a threshold policy with threshold $t$, i.e.,
\begin{multline*}
c_e^{(i)}(t) + c_c^{(i)} + w_i(t) + \mathbb{E}[V_i(t_+)|t,1] \\= c_e^{(i)}(t) + \mathbb{E}[V_i(t_+)|t,0],
\end{multline*}
which yields
\begin{align}
c_c^{(i)} + w_i(t) = &\mathbb{E}[V_i(t_+)|t,0] - \mathbb{E}[V_i(t_+)|t,1] \nonumber\\
=& \lambda_i[V_i(t+1)-V_i(0)].\label{eq: raw whittle index expression}
\end{align}
Under the threshold policy with threshold $t$, the relative value functions $V_i(\cdot)$ should satisfy, for $0\leq t' < t$
\begin{align}
V(t') + \rho_i = c_e^{(i)}(t') + V(t'+1),\label{eq: poisson eq under threshold policy}
\end{align}
where the average cost under the threshold policy is the summation of the estimation errors and communication costs, i.e.,
\begin{align*}
\rho_i = J_e^{(i)}(t) + \frac{1}{\lambda_i t+1}(c_c^{(i)}+w_i(t)),
\end{align*}
and, since transmission and no transmission should have same costs,
\begin{align}
V(t) + \rho_i = c_e^{(i)}(t) + w_i(t) + c_c^{(i)} + V(t+1).\label{eq: poisson eq under threshold policy at threshold}
\end{align}
Plug~\eqref{eq: poisson eq under threshold policy}-\eqref{eq: poisson eq under threshold policy at threshold} in~\eqref{eq: raw whittle index expression} and replace $t$ with $\tau^{(i)}$, we can obtain the expression for $w_i(\tau^{(i)})$ stated in the theorem.

\begin{IEEEbiography}[]{Shuang Wu}
\end{IEEEbiography}

\begin{IEEEbiography}[]{Kemi Ding}
\end{IEEEbiography}

\begin{IEEEbiography}[]{Peng Cheng}
\end{IEEEbiography}

\begin{IEEEbiography}[]{Ling Shi}
\end{IEEEbiography}

\end{document}